\input{aipcheck}

\documentclass[
    ,final            
]
{aipproc}

\layoutstyle{8x11double}

\begin{document}
\title{Photon spectra from quark generation by WIMPs}


\author{J.\,A.\,R.\,Cembranos, A. Cruz-Dombriz$^{1}$,  A.\,Dobado, R.\, Lineros$^{2}$ and A.\,L.\,Maroto
}{
 address={
Departamento de F\'{\i}sica Te\'orica I, Universidad Complutense de Madrid, E-28040 Madrid, Spain.
 \\
$^{1}$ also at Department of Mathematics and Applied Mathematics,
University of Cape Town, 7700 South Africa.
 \\
$^{2}$ IFIC, CSIC-Universitat de Valencia, Ed.\, Instituts, Apt. 22085, 46071 Valencia, Spain.
}
}

\begin{abstract}
%
If the present dark matter (DM) in the Universe annihilates into
Standard Model (SM) particles, it must contribute to the gamma ray fluxes
that are detected on the Earth. 
The magnitude of such contribution
depends on the particular DM candidate, but certain
features of these 
spectra
may be analyzed in a 
model-independent
fashion. 
%
%
In this work we provide the fitting formula valid for the simulated photon spectra from WIMP 
annihilation into light quark-anti quark  ($q\bar{q}$) channels 
in a wide range of WIMP masses. We illustrate our results for the $c\bar{c}$ channel. 
\end{abstract}
\maketitle
\section{I. Introduction}
According to present observations of large scale structures, CMB anisotropies 
and light nuclei abundances, 
DM cannot
be accommodated within the SM of elementary particles.
Indeed, DM presence is a required component on cosmological 
scales, but also to provide a 
satisfactory description of rotational speeds of galaxies,
orbital velocities of galaxies in clusters, gravitational lensing
of background objects by galaxy clusters 
and the temperature distribution of hot gas in galaxies and
clusters of galaxies.
The experimental determination of the DM nature will require the
interplay of collider experiments 
and astrophysical
observations. These searches use to be classified in direct or
indirect searches (see [1] and references in Introduction 
in [2]) 
. Concerning direct ones, the 
elastic scattering of DM particles from nuclei should lead directly 
to observable nuclear recoil signatures although
the weak interactions between DM and the 
standard matter makes DM direct detection extremely difficult.

On the other hand, DM might be detected indirectly, by observing
their annihilation products  into standard model particles. Thus, even if WIMPs
(Weakly Interacting Massive Particles) are stable, two of them may annihilate
into ordinary matter such as quarks
, leptons and gauge bosons. Their annihilation in different places (galactic halo, Sun, etc.)
produce cosmic rays to be discriminated through distinctive
signatures from the background. After WIMPs annihilation a
cascade process occurs. In the end the 
stable particles: neutrinos, gamma rays, antimatter... 
may be observed through different devices. Neutrinos 
and gamma rays have the advantage of maintaining their original
direction due to their null electric charges. 
%

This communication precisely focuses 
on photon production coming 
from $q\bar{q}$ channels (except $t\bar{t}$ channel).
Photon fluxes in specific DM models are usually obtained
by software packages such as DarkSUSY and micrOMEGAs 
based on PYTHIA Monte Carlo event generator [3] 
after having fixed a WIMP mass 
for the particular SUSY model under consideration. 
In this sense, the aim of this investigation
is to provide
fitting functions for the photon
spectra corresponding to each individual annihilation $q\bar{q}$ channel
and, in addition, determine the dependence of such spectra on the
WIMP mass in  a model independent way. This would allow to
apply the results to alternative DM candidates for which
software packages have not been developed.
On the other hand, the information about channel contribution and
mass dependence can be very useful in order to identify gamma-ray
signals for specific WIMP candidates and may also provide
relevant information about the photon energy distribution 
when $q\bar{q}$ pairs annihilate.

Let us remind that the $\gamma$-ray flux from the annihilation 
of two WIMPs of mass $M$ into two SM 
particles coming from all possible annihilation channels (labelled by the subindex $i$) is given by:
\begin{eqnarray}
\label{eq:integrand}
\frac{{\rm d}\,\Phi_{\gamma}^{{\rm DM}}}{{\rm d}E_{\gamma}}&=&\frac{1}{4\pi M^2}
 \sum_i\langle\sigma_i v\rangle
\frac{{\rm d}N_\gamma^i}{{\rm d}E_{\gamma}}
 \nonumber\\
&& \;\, \times \;\,
 \frac{1}{\Delta\Omega} \int_{\Delta\Omega} {\rm d}\Omega
 \int_{\rm l.o.s.} \rho^2[r(s)]\ {\rm d}s\;,
\end{eqnarray}
%
%
where $\langle \sigma_{i} v \rangle$ holds for
the thermal averaged annihilation cross-section of two 
WIMPs into two ($i^{th}$ channel) SM 
particles and $\rho$ is the DM density.
The integral is performed along the line of sight (l.o.s.) to the
target and averaged over the detector solid angle $\Delta\Omega$.
%
%
%

%
%
%
%
\section{III. Procedure}
We have used the particle physics PYTHIA software 
[3] to obtain our results.
The WIMP annihilation is usually 
splited into two separated processes: The first describes the
annihilation of WIMPs and its SM output.
The second one considers the evolution 
of the obtained SM unstable products.
Due to the expected velocity dispersion of DM, we expect
most of the annihilations to happen quasi-statically. This fact 
allows to state that by considering different  
center of mass ($CM$) energies for the obtained SM particles pairs 
from WIMP annihilation process, we are
indeed studying different WIMP masses, i.e. $E_{{\rm CM}} \simeq 2\,M$. 
The procedure to obtain the photon spectra is thus straightforward: For a given pair of 
SM particles which are produced in the WIMP annihilation, we count the number of photons
in bins for the variable $x\equiv E_{\gamma}/M$.
%


Once the PYTHIA simulations 
have been performed, the required parametrization
to fit the data
for the $q\bar{q}$ channels (except $t\bar{t}$)
may be written as:
\begin{eqnarray}
&&\frac{{\rm d}N_{\gamma}}{{\rm d}x}= \frac{a_{1}}{x^{1.5}}
{\rm exp}
\left(-b_{1} x^{n_1}-b_2 x^{n_2} -\frac{c_{1}}{x^{d_1}} 
+\frac{c_2}{x^{d_2}}\right) \nonumber\\
&+& q
{\rm ln}
\left[p(1-x)\right]\frac{x^2-2x+2}{x}
\label{general_formula}
\end{eqnarray}
The parameters in expression \eqref{general_formula}
were considered to be WIMP mass dependent. After a fitting process 
they were determined for
different WIMP masses, in a range varying from 50 to 7000 (or 8000) GeV. 
Mass dependences for the parameters in \eqref{general_formula}
were fitted by using power laws.
\begin{table}
\begin{tabular}{rrrrrr}
\hline
  \tablehead{1}{r}{b}{$M$ (GeV)} 
  & \tablehead{1}{r}{b}{$b_1$}
  & \tablehead{1}{r}{b}{$n_1$}
  & \tablehead{1}{r}{b}{$c_1$}
  & \tablehead{1}{r}{b}{$d_1$}   
  & \tablehead{1}{r}{b}{$p$}  \\
\hline
50   & 5.93 & 2.35 &  0.239 & 0.428 & 210   \\
100  & 5.48 & 2.08 &  0.283 & 0.374 & 379   \\
200  & 4.98 & 1.86 &  0.330 & 0.330 & 673   \\
500  & 4.50 & 1.65 &  0.378 & 0.288 & 1230  \\
1000 & 4.00 & 1.50 &  0.406 & 0.264 & 2110  \\
2000 & 3.70 & 1.35 &  0.432 & 0.245 & 4050  \\
5000 & 3.27 & 1.17 &  0.470 & 0.221 & 8080  \\
8000 & 3.08 & 1.11 &  0.494 & 0.208 & 12000 \\
\hline
\end{tabular}
\caption{
$b_{1}$, $n_{1}$, $c_1$, $d_1$ and $p$ parameters 
corresponding to expression \eqref{general_formula} in the $c\bar{c}$ channel.
Mass independent parameters in \eqref{general_formula}
for this channel are $a_1=5.58$ ; $b_2=7.90$ ; $n_2=0.686$ ; $c_2=0.0$ ;  $q=9.00\cdot10^{-4}$.
}
\label{tab:c quark}
\end{table}
\begin{table}
\begin{tabular}{rrr}
\hline
  \tablehead{1}{r}{b}{Parameter} 
  & \tablehead{1}{r}{b}{Interval (GeV)}
  & \tablehead{1}{r}{b}{Power law(s)}
   \\
\hline
$b_{1}$ & $50 \leq M \leq 8000$ & $9.90\,M^{-0.130}$ \\
$n_{1}$ & $50\leq M \leq 8000$ & $4.14\,M^{-0.148}$  \\
$c_{1}$  & $500\leq M \leq 8000$ & $0.210\,M^{0.0951}$ \\
$d_{1}$ & $50 \leq M\leq 8000$ & $1.50\,M^{-0.632}$ \\ 
 & & + $0.479\,M^{-0.0942}$ \\
$p$ & $200 <M\leq 8000$ & $8.11\,M^{0.812}$ \\
\hline
\end{tabular}
\caption{Fitting power laws 
in $c\bar{c}$ channel. 
}
\label{tab:c:quark:2}
\end{table}
\section{III. $c\bar{c}$ channel}
In order to illustrate the explained procedure, we study
the $c\bar{c}$ channel. For this channel there are five
mass dependent parameters in expression \eqref{general_formula}: 
$b_1$, $n_1$, $c_1$, $d_1$ and $p$ presented in Table I.
The mass independent parameters are
$a_1$, $b_2$, $n_2$, $c_2$
($d_2$ is thus irrelevant) and $q$. 
%
In Table II we present the fitting power laws for mass dependent parameters. 
Figure 1 presents spectra for four different WIMP masses whereas  Figure 2
shows fitting power laws for two mass dependent parameters.
The results for the other $q\bar{q}$ channels [2] are completely analogous even 
though for each channel, the parameters which are mass dependent may be 
different to the ones for the $c\bar{c}$ channel. 
\begin{figure}
\includegraphics[height=.25\textheight]{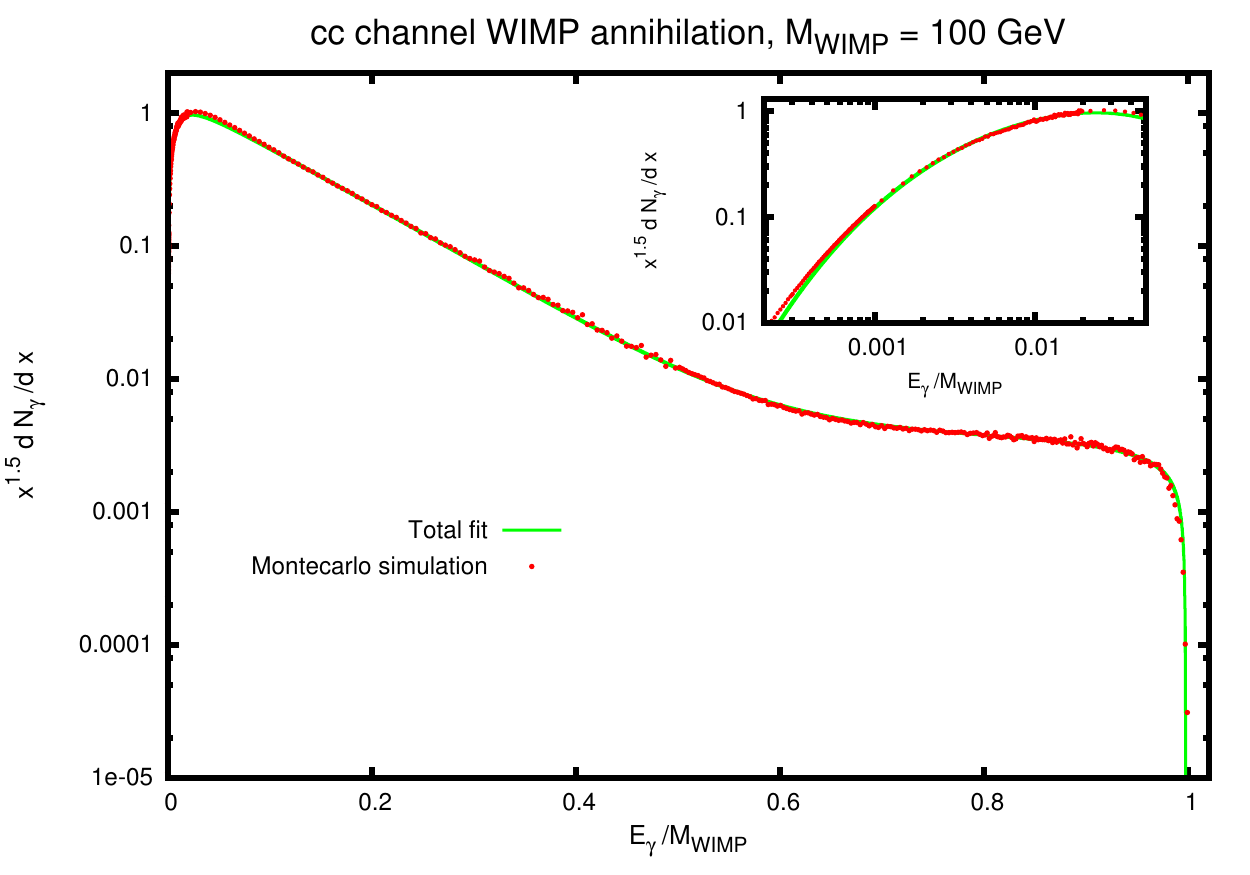}
\includegraphics[height=.25\textheight]{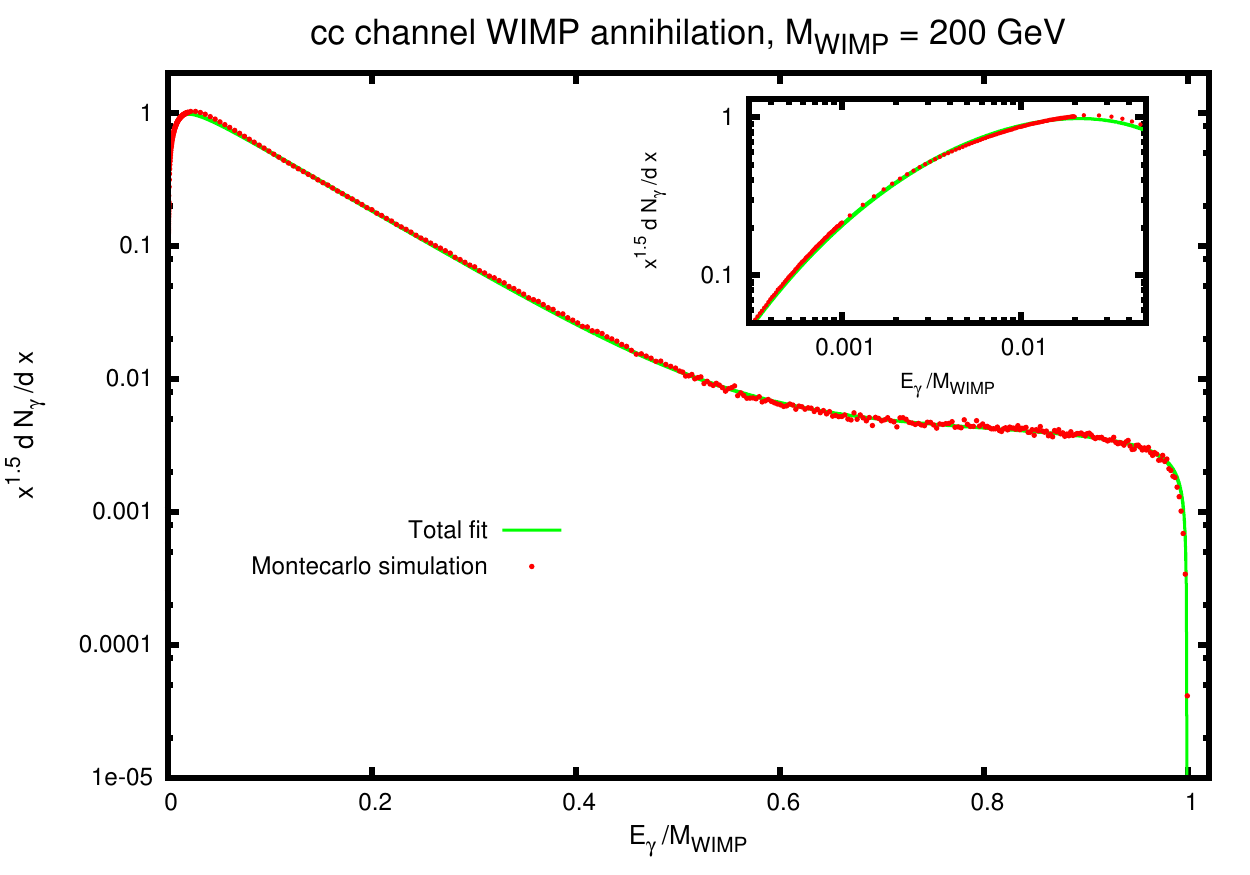}
\end{figure}

\begin{figure}
\includegraphics[height=.25\textheight]{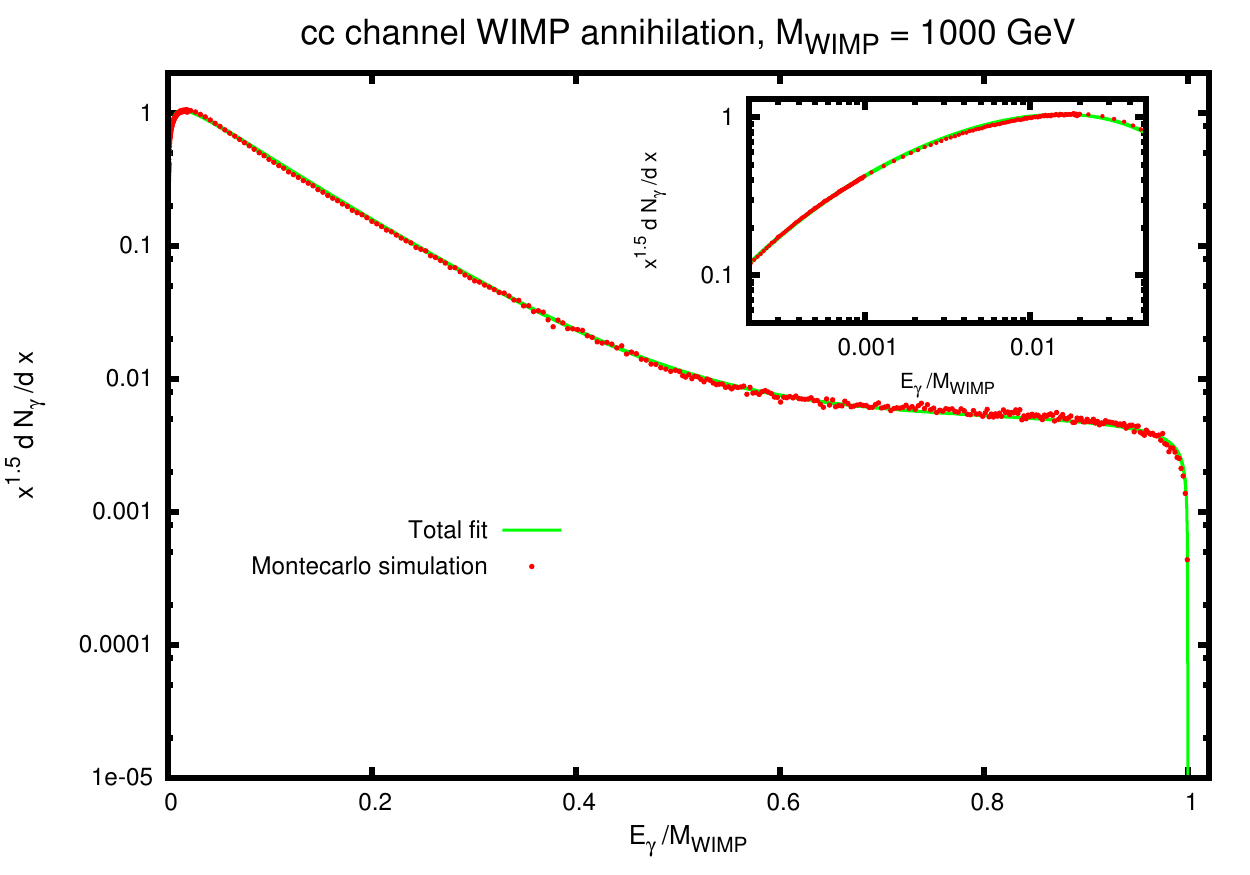}
\includegraphics[height=.25\textheight]{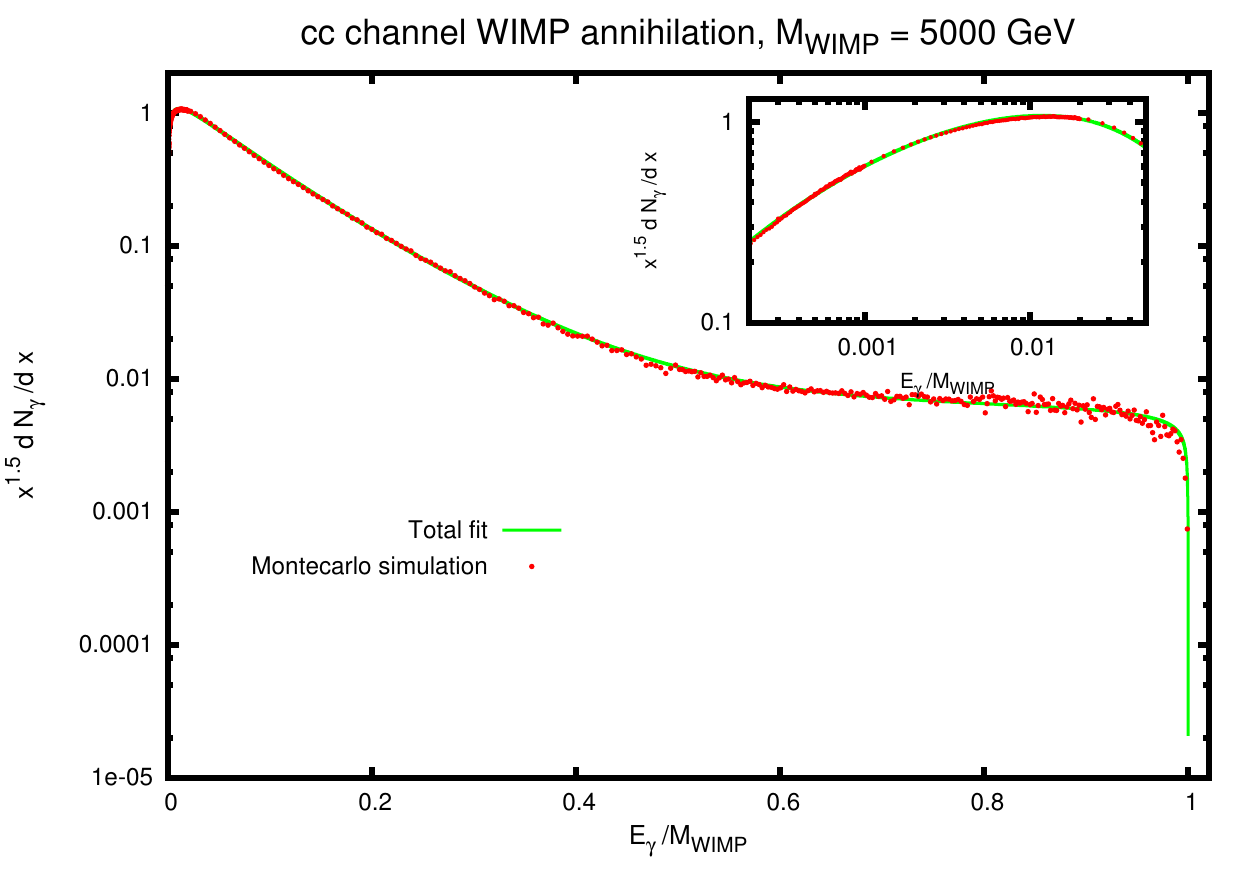}
\caption{Photon spectra for four different WIMP masses (100, 200, 1000 and 5000 GeV) in the $c\bar{c}$ 
annihilation channel. Red dotted points are PYTHIA simulations and solid lines correspond to the proposed fitting functions.}
\end{figure}
\begin{figure}
\includegraphics[height=.25\textheight]{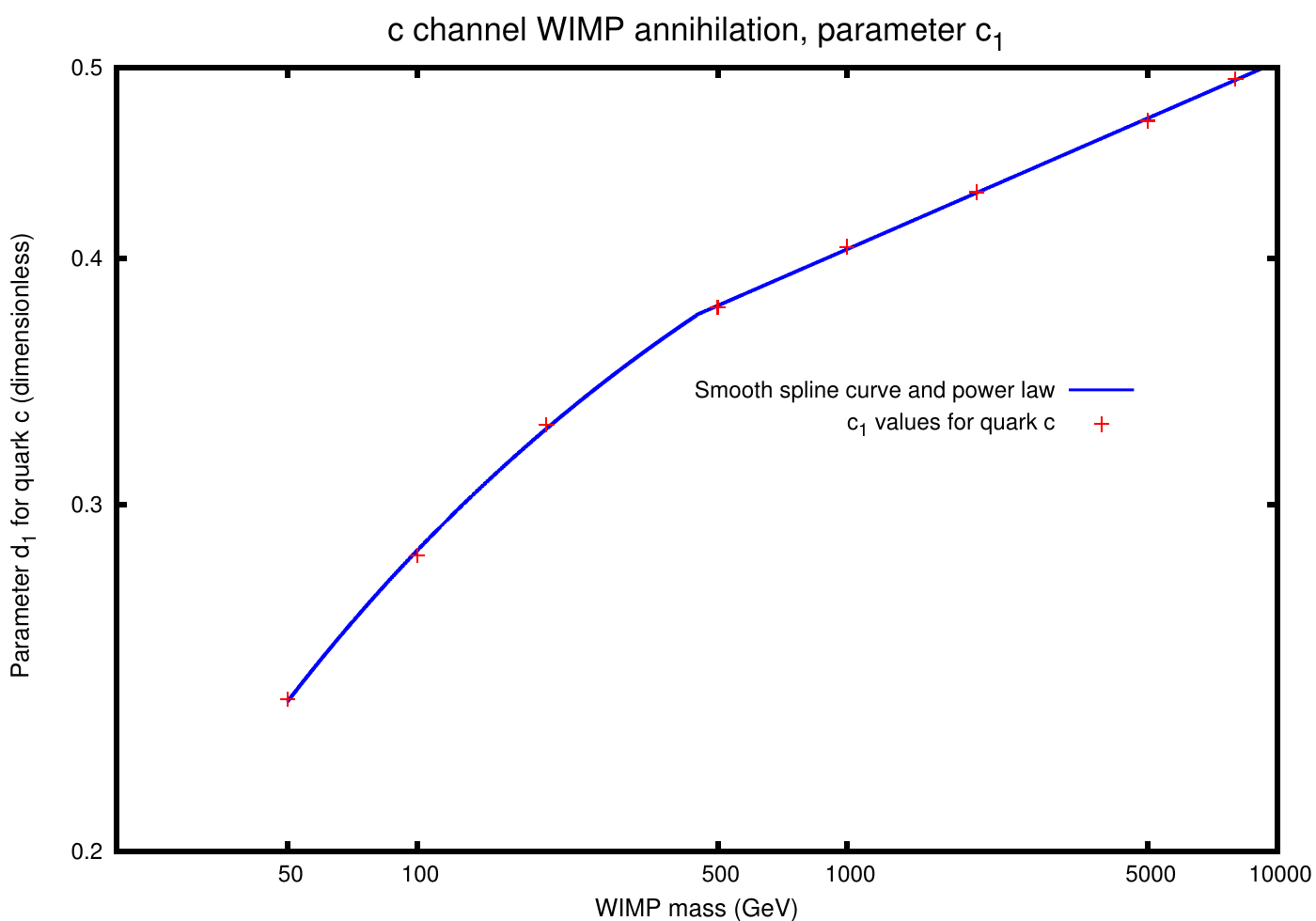}
\includegraphics[height=.25\textheight]{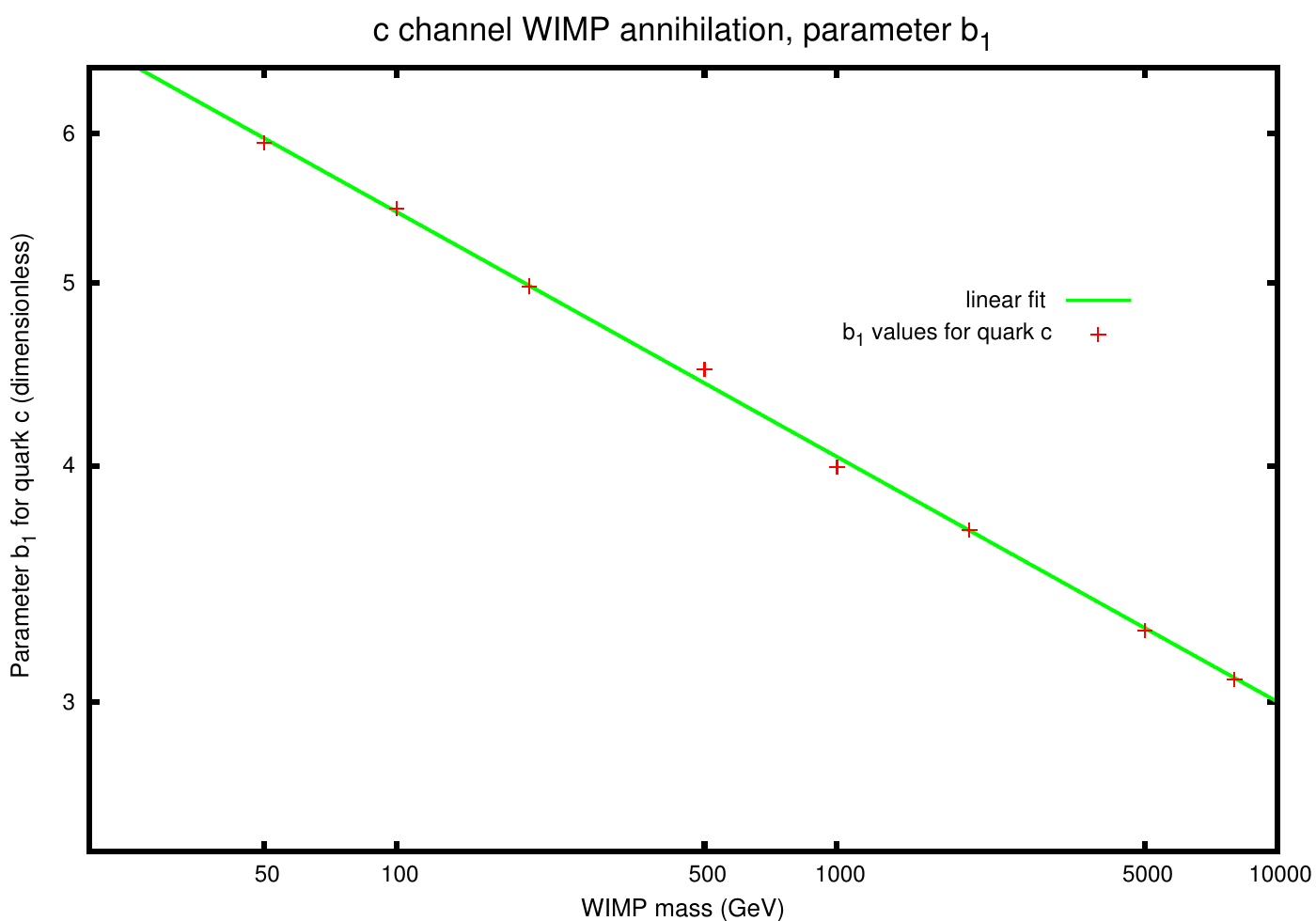}
\caption{Mass dependence of  $c_1$ and $b_1$ parameters for $c\bar{c}$ annihilation channel. Crossed points 
are parameters values found after the fitting process for each WIMP mass and solid lines correspond to the proposed fitting functions.}
\end{figure}
\section{IV. Conclusions}
In this work, we have studied the photon spectra coming from
WIMP pair annihilation into $q\bar{q}$
pairs for all the channels (except $t\bar{t}$). 
The covered WIMP mass range was from 50 GeV to 8000 GeV. Simulated spectra 
covered the whole accessible energy interval: from extremely low energetic photons till
photons with one half of the available total center of mass energy.

Once the spectra were simulated, an 
analytical expression \eqref{general_formula} was proposed
to fit the data. This expression depends on  
both WIMP mass dependent and independent
parameters. 
Our results can 
both provide a better understanding of the DM annihilation channels into photons
and save an important amount of unnecessary Monte Carlo simulations.
This fact is
particularly important for high energy photons, whose 
production rate is very suppressed.
%
%
%
%
%
%
%
%


Calculations for all $q\bar{q}$ channels are available at the website
\url{http://teorica.fis.ucm.es/}$\sim$\url{PaginaWeb/downloads.html}. 
This work was partially supported by 
MULTIDARK CSD2009-00064.
%
%
\bibliographystyle{aipproc}  

\end{document}